\pdfoutput=1 
\documentclass{JINST}
\usepackage[caption=false]{subfig}
\usepackage{array}
\newcolumntype{x}[1]{>{\centering\let\newline\\\arraybackslash\hspace{0pt}}p{#1}}
\usepackage{amsmath}

\title{Characterization and Modeling of a Water-based Liquid Scintillator}
\author{L. J. Bignell$^a$\thanks{Corresponding author.}~,
        D. Beznosko$^{a,c}$,
        M. V. Diwan$^a$,
        S. Hans$^b$,
        D. E. Jaffe$^a$,
        S. Kettell$^a$,
        R. Rosero$^b$,
        H. W. Themann$^{a,d}$,
        B. Viren$^a$,
        E. Worcester$^a$,
        M. Yeh$^b$,
        and C. Zhang$^a$\\
\llap{$^a$}Physics Department, Brookhaven National Laboratory,\\ Upton NY, USA\\
\llap{$^b$}Chemistry Department, Brookhaven National Laboratory,\\ Upton NY, USA\\
\llap{$^c$}Now at Department of Physics, Nazarbayev University,\\ Astana , Kazakhstan\\
\llap{$^d$}Now at Center for Axion and Precision Physics Research, Institute for Basic Science,\\ Daejeon, Republic of Korea\\
E-mail: \email{lbignell@bnl.gov}}

\abstract{
We have characterised Water-based Liquid Scintillator (WbLS) using low energy protons, UV-VIS absorbance, and fluorescence spectroscopy. We have also developed and validated a simulation model that describes the behaviour of WbLS in our detector configurations for proton beam energies of 210 MeV, 475 MeV, and 2 GeV and for two WbLS compositions. Our results have enabled us to estimate the light yield and ionisation quenching of WbLS, as well as to understand the influence of the wavelength shifting of Cherenkov light on our measurements. These results are relevant to the suitability of WbLS materials for next generation intensity frontier experiments.
}


\begin{document}

\section{Introduction}\label{sec:Intro}

Water-based liquid scintillator (WbLS) is a recently developed scintillating material that has been identified as a candidate detector medium for the next generation of intensity frontier particle physics experiments \cite{Yeh2011}. WbLS is a stable scintillating emulsion with a large fraction of aqueous phase.

\begin{figure}[tbp]
\centering
\includegraphics[width=0.8\columnwidth]{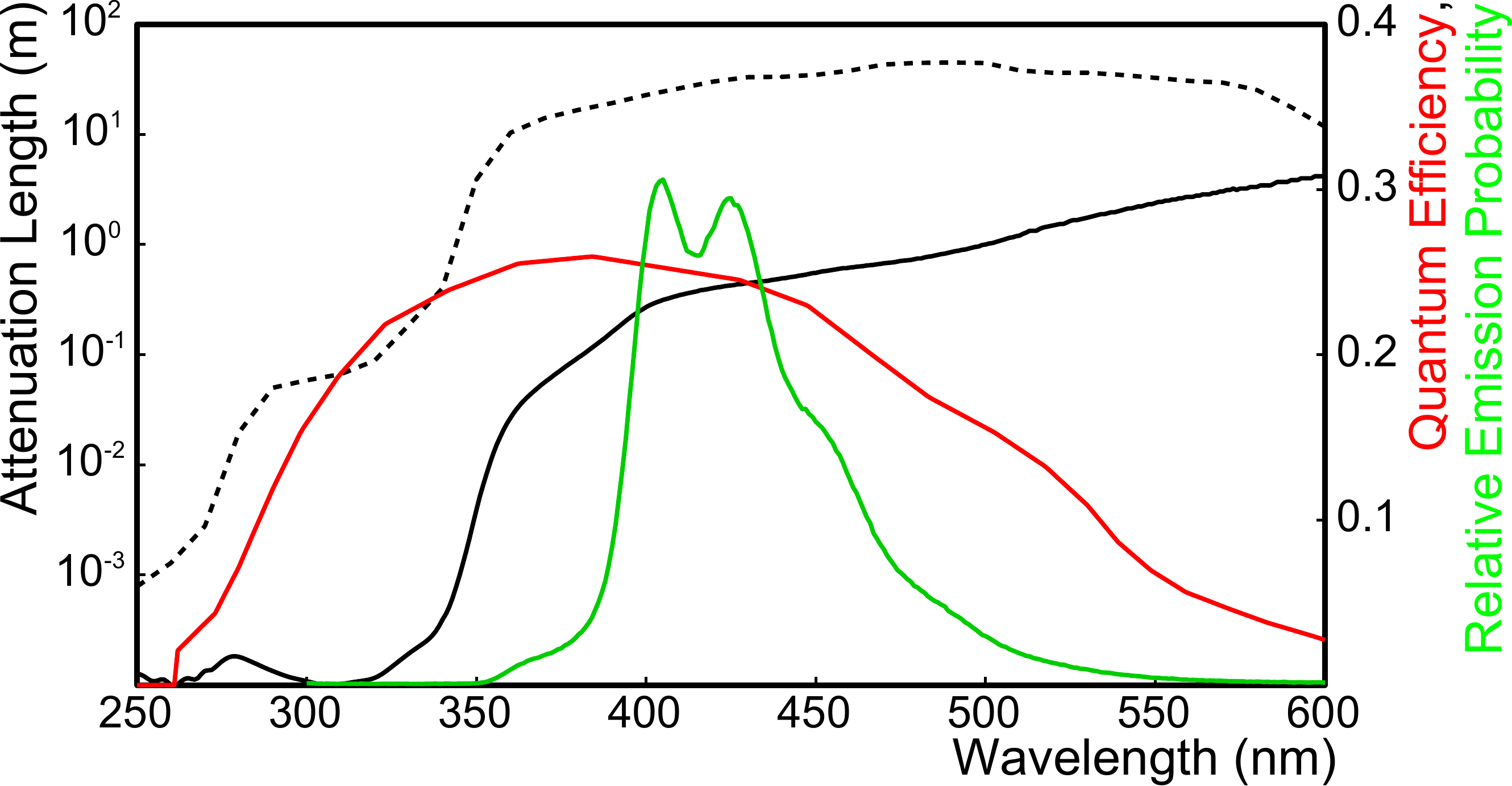}
\caption{The optical attenuation spectra for the 1\% WbLS used in this study (solid black) and for a more recent formulation of 1\% WbLS (dashed black). The 1\% WbLS fluorescence spectrum (green) and photomultiplier tube (Hamamatsu R7723) quantum efficiency (red) are also indicated. The fluorescence spectrum was measured using an exciting wavelength of 290 nm.}
\label{fig:UVVIS}
\end{figure}

With suitable chemical purification and material optimisation, WbLS may be produced with an optical attenuation of tens of meters in the photon energy range relevant to detection by bialkali photomultiplier tubes (PMTs) (figure \ref{fig:UVVIS}). Such a long optical attenuation length would make Cherenkov imaging feasible in a large WbLS detector, which is not possible in large detectors filled with ordinary liquid scintillator. As a scintillating material, WbLS is sensitive to particle interactions below the Cherenkov threshold.

The aqueous phase of WbLS may also allow the loading of metallic ions that are not easily incorporated into the non-polar organic solvents that are typical of liquid scintillators. This may permit novel schemes to incorporate metal ions at high concentrations in the WbLS material. We are currently evaluating the possibility to incorporate candidate isotopes for neutrinoless double beta decay, neutron absorbing isotopes, and high Z materials for gamma attenuation.

The optical and chemical properties of WbLS suggest that it may be an effective detection medium to be used in a particle detector capable of a broad range of measurements. Neutrinoless double beta decay, high energy neutrino beam measurements, proton decay, and low energy neutrino physics have been proposed as plausible measurements using a large scale (30 -- 100 kiloton) WbLS detector \cite{Alonzo2014etal}.

In order to rigorously evaluate the suitability of WbLS for such an application, a comprehensive understanding of the performance of the material must be developed. Basic material properties such as the light yield, the optical attenuation length, and the emission spectrum must be measured. A model of the response of the WbLS to ionising radiation and optical and ultraviolet photons must also be developed and validated.

In this study, we have performed several experimental tests aimed at elucidating these properties, and we have also developed a simulation model that successfully predicts the performance of the WbLS material in our measurement geometries. It is our intention to further develop and validate this simulation model using larger volume geometries to permit accurate study of the performance of a proposed large experiment that uses WbLS as its detection medium.

\section{Experiment}\label{sec:Expt}
\subsection{WbLS Samples}
WbLS samples composed of 0.4\% (WbLS-1) and 1\% (WbLS-2) liquid scintillator by mass were prepared, as well as pure liquid scintillator (LS) samples. The pure LS was identical to that used by the Daya Bay experiment \cite{An2012etal}. Table \ref{tab:ScintComp} details the compositions of these samples.

\begin{table*}[tbp]
\caption{The chemical composition of the scintillators used in this study. The organic phase is given as a mass fraction of the final solution. The fluor concentrations are given relative to the organic solvent. Chemical names are abbreviated as follows: PC = Pseudocumene, PPO = 2,5- diphenyloxazole, bis-MSB = 1,4- Bis (2-methylstyryl) benzene, LAB = Linear Alkyl Benzene, LS = Liquid Scintillator.}
\smallskip
\centering
\begin{tabular}{|cccc|}
\hline
Sample Name & Organic Phase & Primary Fluor & Secondary Fluor \\
\hline
WbLS-1  & PC, 0.4\%     & PPO, 0.4 g/L  & bis-MSB, 3 mg/L  \\
WbLS-2  & PC, 0.99\%    & PPO, 1.36 g/L & bis-MSB, 7.48 mg/L \\
LS      & LAB, 100\%    & PPO, 2 g/L    & bis-MSB, 15 mg/L \\
\hline
\end{tabular}
\label{tab:ScintComp}
\end{table*}

\subsection{Optical Characterization Measurements}
Ultraviolet-visible 
absorption spectroscopy of water, WbLS-1, and WbLS-2 were taken using a Shimandzu UV-1800 spectrophotometer. 
The optical attenuation spectrum of WbLS-2 is shown in figure \ref{fig:UVVIS}, along with that of a more recent WbLS formulation that also consists of 1\% scintillator, the fluorescence emission spectrum, and the photomultiplier quantum efficiency spectrum for the PMTs used in this study. The optical attenuation of the more recent formulation of a LAB-based WbLS is markedly less than WbLS-2, after the purification of all starting materials and the addition of a surface-scattering reduction agent.

\begin{figure}[tbp]
\centering
\includegraphics[width=0.8\columnwidth]{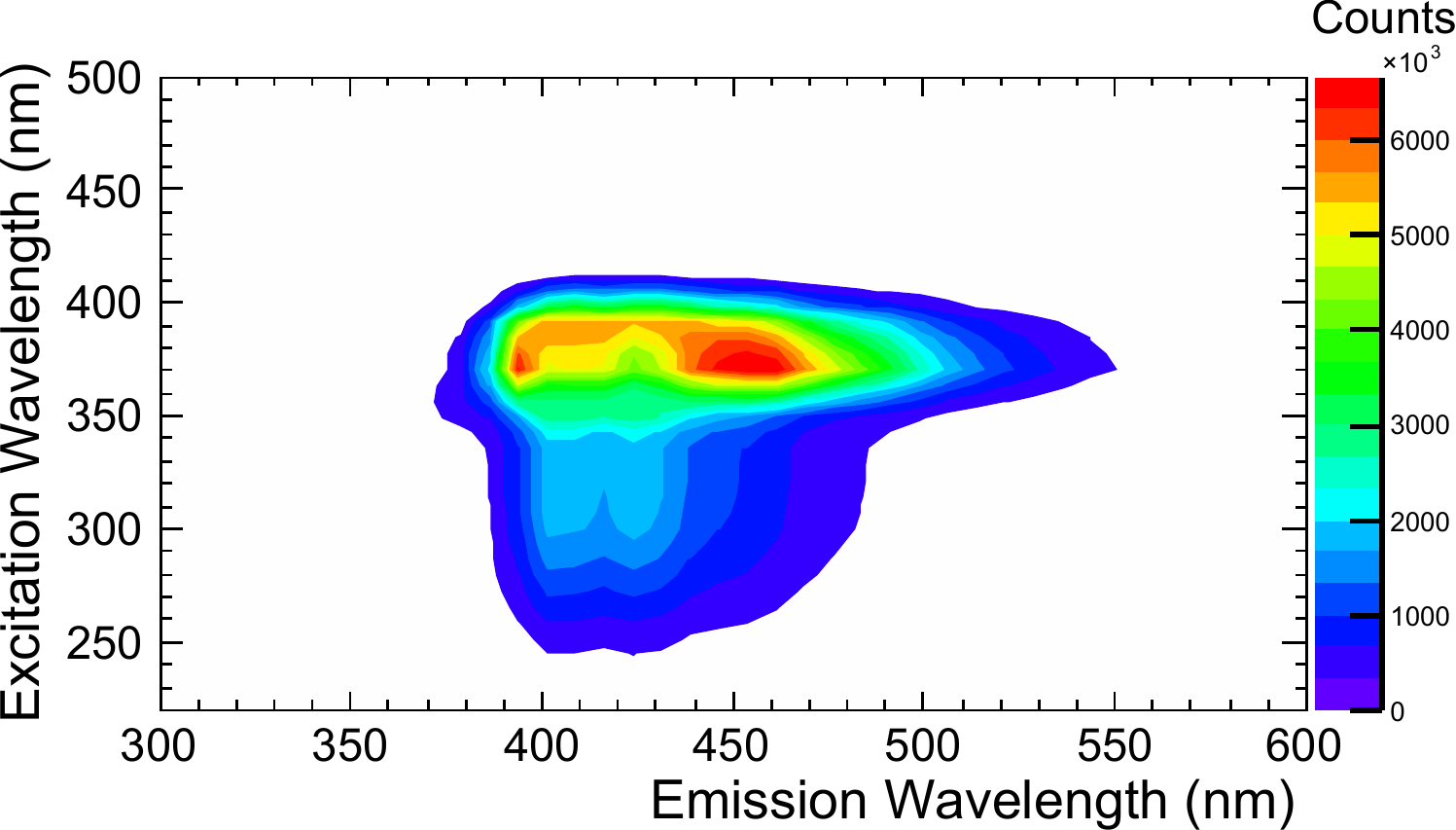}
\caption{Excitation-emission spectrum for WbLS-2.}
\label{fig:ExEmMap}
\end{figure}

Fluorescence spectroscopy was carried out using a PTI fluoresence spectrometer. The excitation-emission spectrum of WbLS-2 is shown in figure \ref{fig:ExEmMap}, which exhibits an excitation wavelength dependence of the emission spectrum shape.

\subsection{Proton Beam Measurements}
\label{sec:BeamInfo}
\begin{figure}[tbp]

\centering
\includegraphics[width=0.8\columnwidth]{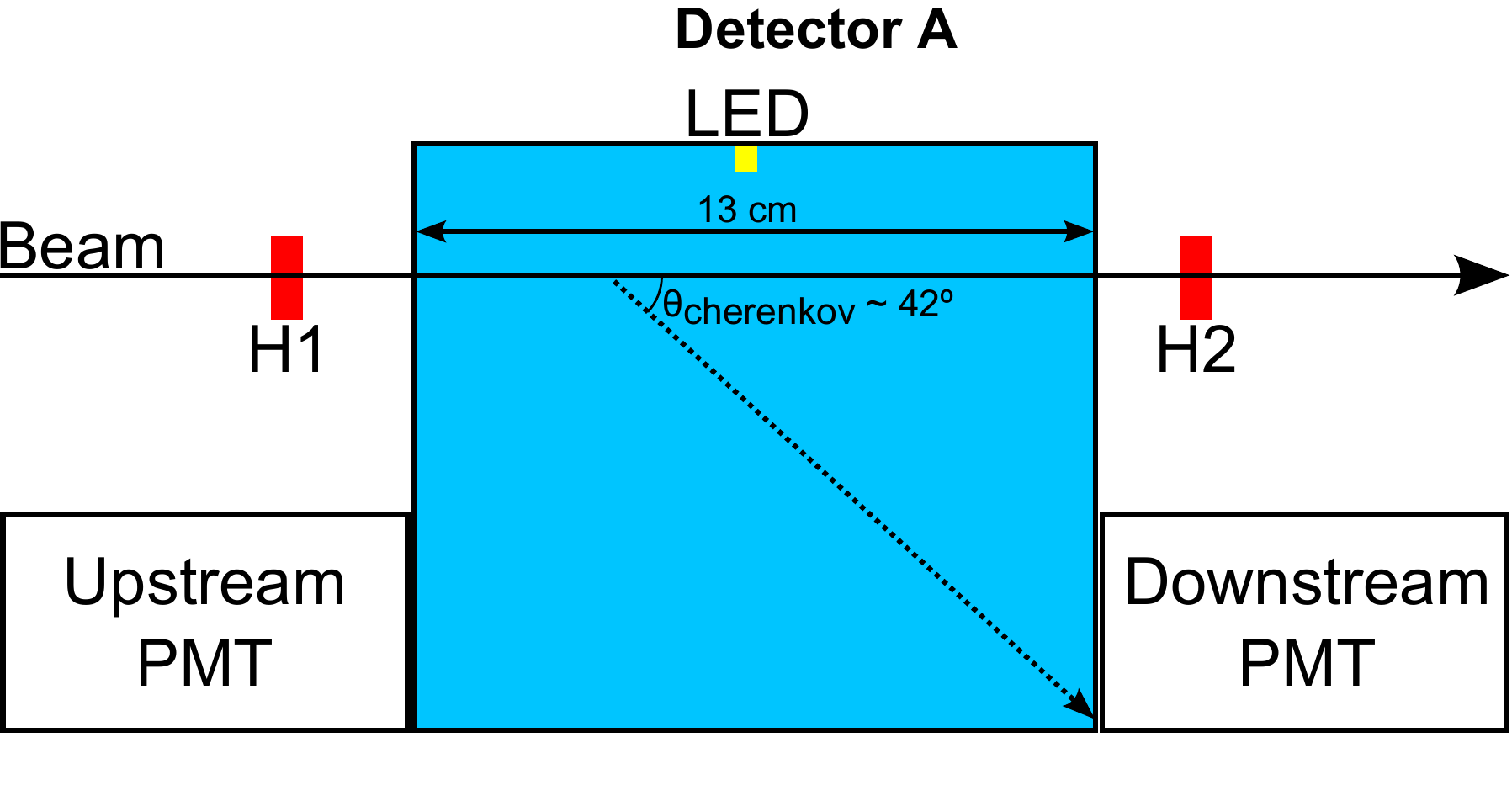}

\includegraphics[width=0.8\columnwidth]{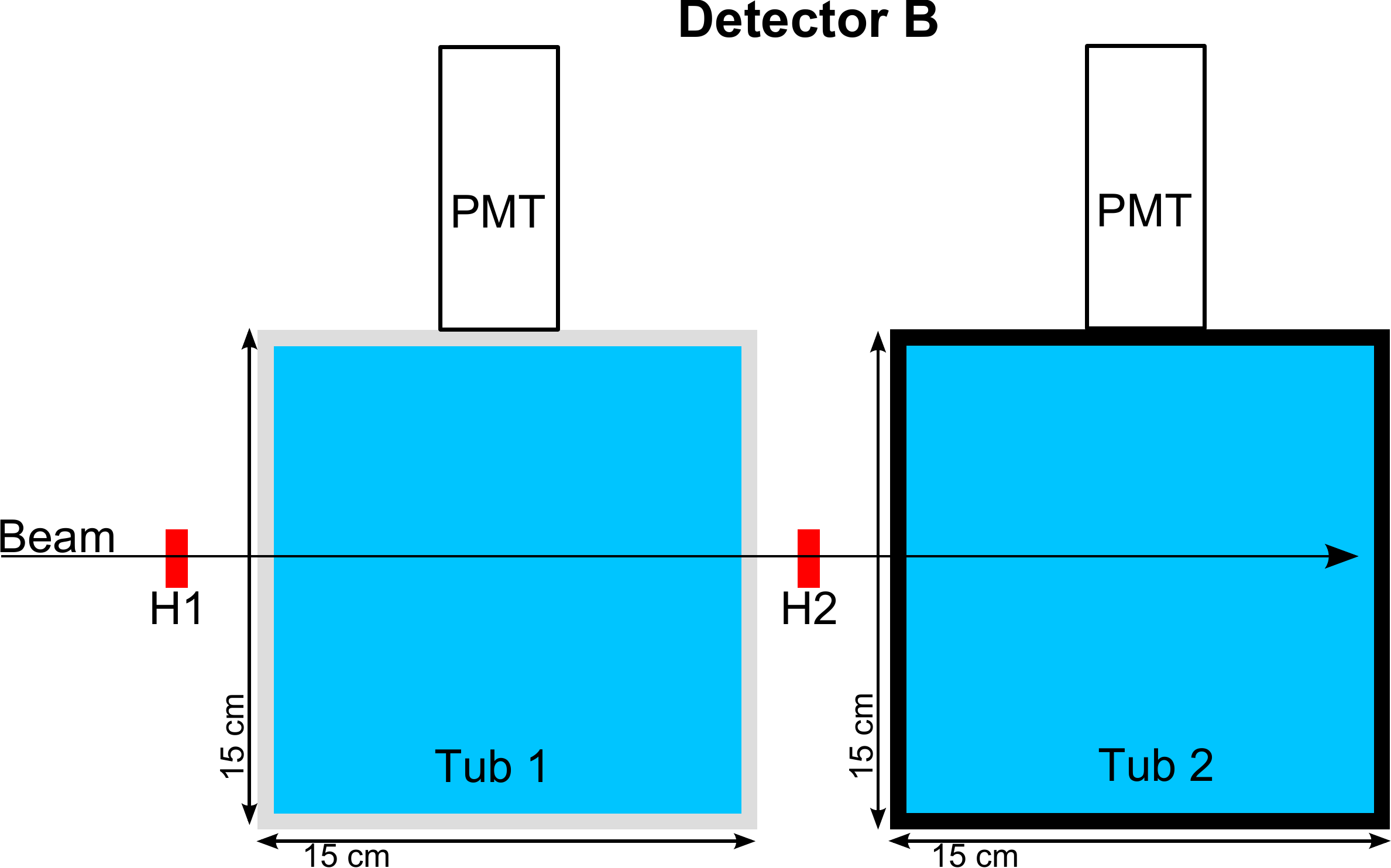}
\caption{A schematic drawing of the instruments used in the proton beam measurements, Detector A 
and Detector B. 
See the text for details.}
\label{fig:NSRLExpts}
\end{figure}

Two beamline instruments were developed for this study, Detectors A and B (figure \ref{fig:NSRLExpts}). 

\begin{figure}[tbp]
\centering
\includegraphics[width=0.8\columnwidth]{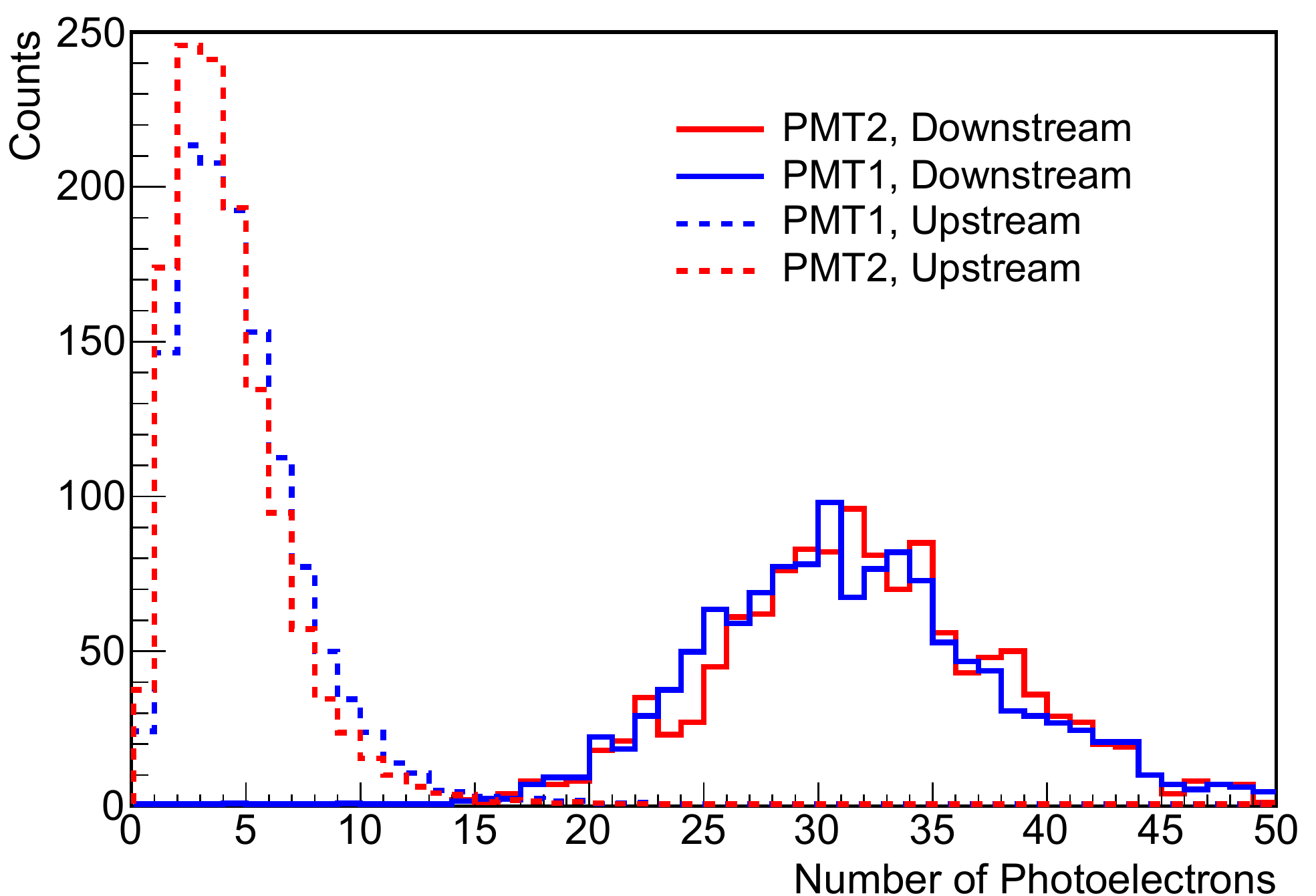}
\caption{The measured photoelectron distributions for identical measurements using the two different photomultipliers (represented as the red and blue traces) in Detector A. The solid traces represent measurements taken downstream of the 2 GeV proton beam on a water target, and the dashed traces represent measurements taken upstream of the 475 MeV proton beam on a WbLS-2 target. The photomultiplier responses are similar for small and large numbers of photoelectrons.}
\label{fig:PMTresponse}
\end{figure}

Detector A was fabricated using black ABS polymer to supress light reflections from the walls. The 130 mm x 115 mm x 63.5 mm detector volume was filled with water during initial irradiations, then WbLS-2. Two PMTs (Hamamatsu R7723, 51 mm diameter) observed the liquid volume; one placed downstream from the beam so as to allow detection of Cherenkov light, and one placed upstream from the beam so as to suppress detection of Cherenkov light. To allow for any variability in the photomultiplier response, all measurements were repeated with the PMT locations relative to the beam direction swapped. The PMT response was found to be similar for each tube (figure \ref{fig:PMTresponse}). We report the averaged values of the PMT response hereafter. Two 2x2x0.5 cm plastic scintillator hodoscopes (H1 and H2) were used for triggering and to define the beam location.

Detector B consists of two right cylindrical 150 mm x 150 mm tubs. Tub 1 was fabricated using 6.35 mm thick white teflon and Tub 2 was fabricated using 6.35 mm thick aluminium coated with black perfluoroalkoxy alkane paint on its inner surface. The tubs were exposed to the proton beam for consecutive fillings of water, WbLS-1, WbLS-2, and LS. Two 2x2x0.5 cm plastic scintillator hodoscopes were used to define the beam.

A 400 nm LED was placed just above each scintillating volume in Detector A to permit an in-situ single photoelectron PMT calibration. The low energy water irradiations were used for single photoelectron PMT calibrations in Detector B.

All samples were investigated using proton beams with incident energies of 475 MeV and 2 GeV at the NASA Space Radation Laboratory at Brookhaven National Laboratory. A 210 MeV incident proton beam was additionally used to investigate all samples in Detector B. 
The beam intensity at all energies was made low enough that the probability of having more than a single proton incident per accelerator bunch was small. An analysis of our results where the two-proton events are separable from the 1 proton events indicates that the fraction of two-proton events varied as a function of beam energy and was less than 10\% in all irradiations. We have estimated the influence of the two-proton events by resampling the simulated single proton distributions presented below. Double proton events have no effect on the results with a large (>10) mean number of detected photoelectrons. The effect on the remaining data is estimated to be <10\% and is included in the systematic uncertainty discussed in section \ref{sec:Results}.

\begin{table}[tbp]
\caption{A summary of the proton beam exposures.}
\smallskip
\centering
\begin{tabular}{|ccc|}
\hline
Instrument & Sample & Incident Proton Energy \\
\hline
Detector A  &   Water   & 2 GeV and 475 MeV             \\
Detector A  &   WbLS-2  & 2 GeV and 475 MeV             \\
Detector B  &   Water   & 210 MeV, 475 MeV, and 2 GeV   \\
Detector B  &   WbLS-1  & 210 MeV, 475 MeV, and 2 GeV   \\
Detector B  &   WbLS-2  & 210 MeV, 475 MeV, and 2 GeV   \\
Detector B  &   LS      & 210 MeV, 475 MeV, and 2 GeV   \\
\hline
\end{tabular}
\label{tab:BeamMeas}
\end{table}

The data readout in all measurements was achieved using a CAEN V1729A 14 bit waveform digitizer sampling at 1 gigasamples per second. The raw waveforms were stored for offline analysis. Data acquisition was triggered by coincident hodoscope events during the accelerator beam gate.

\section{Analysis}

\subsection{Signal Processing Algorithm}

\begin{figure*}[tbp]
\centering
\includegraphics[width=0.8\textwidth]{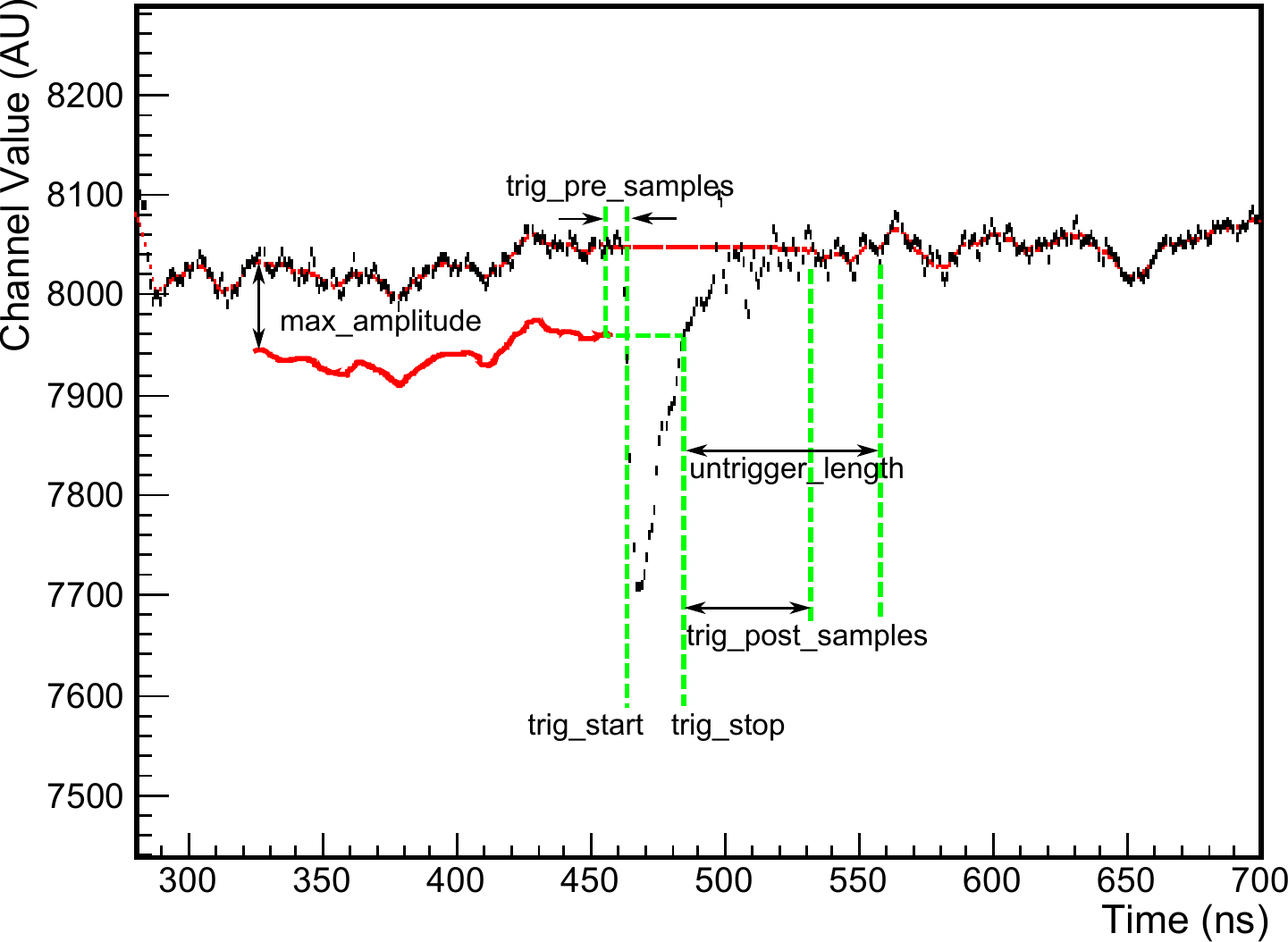}
\caption{A schematic of the waveform data processing algorithm used to determine the pulse information for analysis. The dashed red line is the baseline determined by the algorithm. See the text for details.}
\label{fig:Algo}
\end{figure*}

Signal processing of the acquired waveforms was achieved using a custom algorithm which was developed for background subtraction, timing, and pulse area determination. The algorithm is based upon that used in the DarkSide experiment \cite{Alexander2013etal} and its operation is illustrated in Figure \ref{fig:Algo}.

A moving average of the waveform data is taken as the baseline. A data point is considered to be part of the baseline if the difference between that data point value and the value of the averaged baseline as it was \texttt{trig\_pre\_samples} data points ago is less than the algorithm's threshold parameter, \texttt{max\_amplitude}. If a trigger event occurs (\texttt{trig\_start}), the algorithm waits until the waveform falls below the threshold level, then waits for an additional pile-up pulse for \texttt{untrigger\_length} samples. If a pile-up event occurs, \texttt{trig\_stop} is unset and the piled-up event is considered to be part of the same trigger. Otherwise, the algorithm linearly interpolates between the baseline values at the start and end of the pulse and continues to average the baseline and search for a new trigger. The pulse charge was measured as the summed pulse area between \texttt{trig\_pre\_samples} and \texttt{trig\_post\_samples}, less the baseline. We used \texttt{trig\_pre\_samples} = 3 ns and \texttt{trig\_post\_samples} = 10 ns.

The use of a moving average of the baseline attenuates high frequency components of the signal pedestal noise; allowing a \texttt{max\_amplitude} value of approximately 0.3 photoelectrons.

\subsection{Simulation Model}
We have used Geant4 v10.0 \cite{Agostinelli2003etal} to simulate detectors A and B, the energy deposit in the hodoscopes and scintillator, and the optical photon measurements by the PMTs.

For modeling the optical photon processes, the optical properties of the detector materials were required. Where possible we used measured values. Our measurements of the excitation and emission spectra and optical attenuation coefficient spectrum of the scintillators used in this study were incorporated into the model. The wavelength-shifting absorption coefficient was taken to be equal to the difference between the WbLS optical attenuation coefficient and the optical attenuation coefficient of water. The scintillation emission spectrum was assumed to be identical to that produced when excited by 290 nm photons. The refractive index of water and optical absorbance of water were taken from Segelstein \cite{Segelstein1981}. The refractive index and re-emission probability of liquid scintillator were taken from the Daya Bay simulation model \cite{Wang2009}. The refractive index of WbLS samples was calculated as a linear combination of water and pure liquid scintillator, according to the fraction of scintillating solvent present in the WbLS. The spectral dependence of the WbLS re-emission probability was assumed to be identical to that of pure liquid scintillator, although the amplitude of the re-emission probability was left as a free parameter whilst being constrained to a maximum value of 1. The 2 GeV measurements in Detector A were employed for this optimisation. We assumed that the re-emission probabilities of WbLS-1 and WbLS-2 are identical. The optical absorbance and refractive index data for the UV-transparent acrylic windows were taken from Band \textit{et al.} \cite{Band2012}. The refractive index of the borosilicate glass PMT windows was calculated using the empirical Sellmeier equation parameters provided by Schott \cite{Schott2012}. 
The quantum efficiency spectrum was estimated using data provided by Hamamatsu. 

Several optical parameters were either unknown or poorly known, so that their values were assumed or left as free parameters. The black vessels were assumed to be perfectly unreflective, and the white vessel was assumed to be a wavelength-independent diffuse reflector with the reflectance allowed to vary. The water measurements were used to calibrate other model parameters to reproduce the measured results. The free parameters were the PMT photocathode radius for Detector A and the beam height relative to the centre of the detector vessel for Detector B. The beam height in the Detector B was estimated to be located at the centre of the tub with an uncertainty of $\pm$ 1 cm. The optimal beam height was found to be 0.69 $\pm$ 0.05 cm above the centre of the vessel. The optimal photocathode radius of 22 $\pm$ 1 mm is in fair agreement with the manufacturer's specification of $>$23 mm.

\subsubsection{Wavelength Shifting Model}
For ordinary liquid scintillator with a scintillation yield of $10^{4}$ photons per MeV, the number of scintillation photons produced per centimeter for a minimum ionising particle experiencing a stopping power of 2 MeV/cm is $2 \times 10^{4}$. This vastly outnumbers the yield of Cherenkov photons, which is $\approx 600$ photons per centimeter in water integrated over the 200-500 nm spectral range. As more than half of these photons fall in the 200 to 300 nm spectral range, the Cherenkov light accounts $<$1.5\% of the detectable light from a minimum ionising particle in ordinary liquid scintillator. WbLS has a much lower scintillation light yield than ordinary liquid scintillator, so that the number of Cherenkov photons may in fact be greater than the number of scintillation photons. This Cherenkov light may be directly detected or absorbed by the WbLS. Wavelength-shifting (WLS) of the absorbed light may occur with a probability described by the WbLS's re-emission probability. Understanding the Cherenkov and scintillation light generation and WLS is therefore an important aspect of evaluating the suitability of WbLS for large detector geometries.

\begin{figure}[tbp]
\centering
\includegraphics[width=0.8\columnwidth]{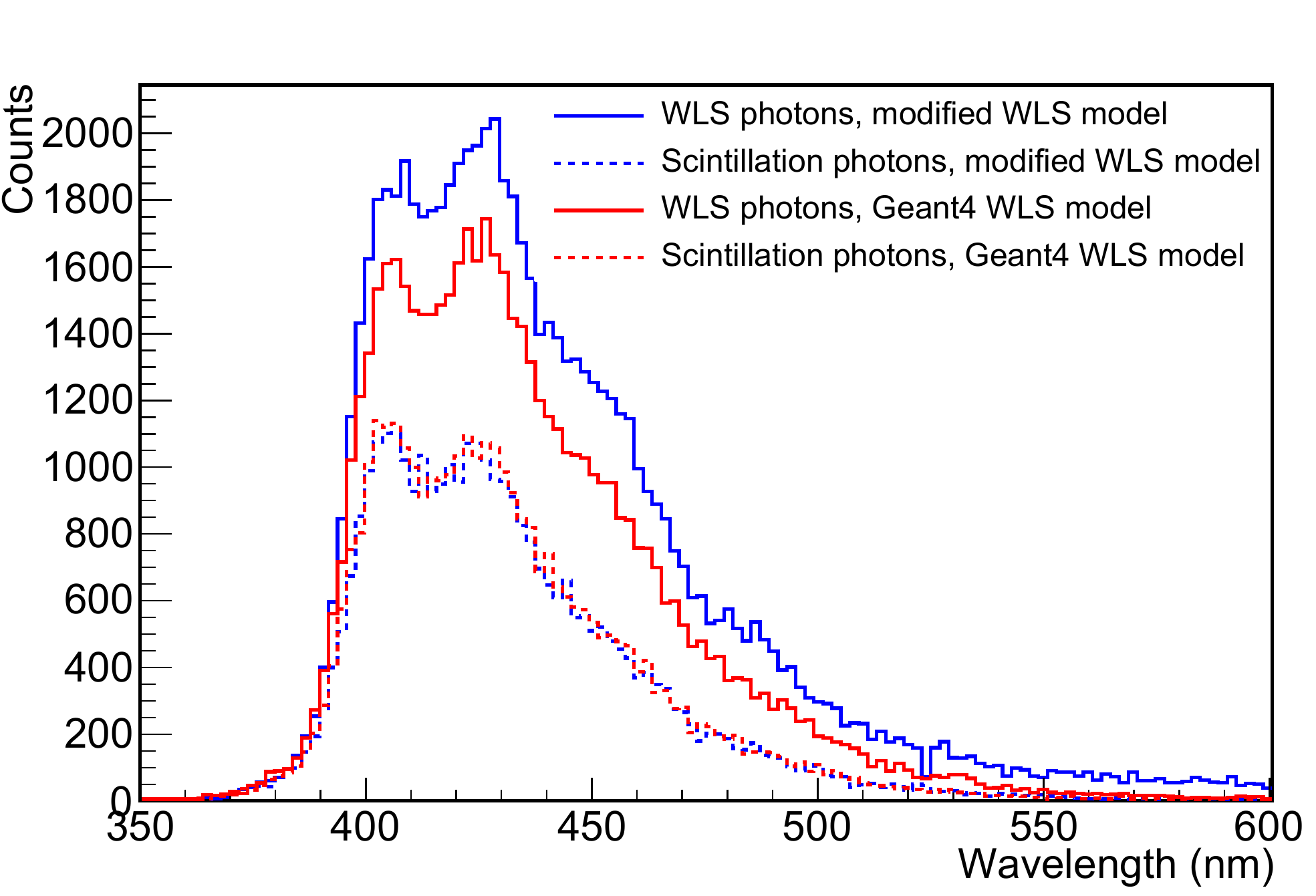}
\caption{The predicted optical photon spectrum in Detector A due to the scintillation (dashed traces) and wavelength-shifting (solid traces) processes, as measured by the upstream PMT for incident 2 GeV protons. The original Geant4 wavelength-shifting model results are represented using red traces, and the results from the modified wavelength-shifting model developed in this work are represented using blue traces.}
\label{fig:WLSmodels}
\end{figure}

This sensitivity to WLS of Cherenkov photons has motivated us to extend the Geant4 optical WLS model. Our modifications to the WLS model have allowed both the re-emission probability and the emission spectrum to depend upon the wavelength of the absorbed photon. Figure \ref{fig:WLSmodels} compares the model predictions of the optical photon spectrum incident on the upstream PMT in Detector A due to a beam of 2 GeV protons, when it is filled with WbLS-2. Although a large proportion of wavelength-shifted light is predicted by both models, the original Geant4 model predicts 21\% less WLS light. There is also some spectral distortion evident in the modified WLS model spectrum at around 450 nm due to the influence of the changes in the fluorescence emission spectrum that occur mainly for exciting wavelengths between about 350 nm and 400 nm (figure \ref{fig:ExEmMap}). Our modified WLS model predicts a mean number of detected photons that is 17\% greater than the Geant4 WLS model.  

\subsubsection{Scintillation Parameter Optimisation}
The scintillation light yields and ionisation quenching factors of all samples were taken as free model parameters. The light yields were estimated by performing a $\chi^2$ optimisation of the simulated photoelecton distribution to the measurement, using data from Detector B when irradiated by 475 MeV protons. The ionisation quenching factor optimisation was performed in a similar manner, using Detector B's 210 MeV proton irradiation.

\begin{figure*}[tbp]
\centering
\includegraphics[width=\textwidth]{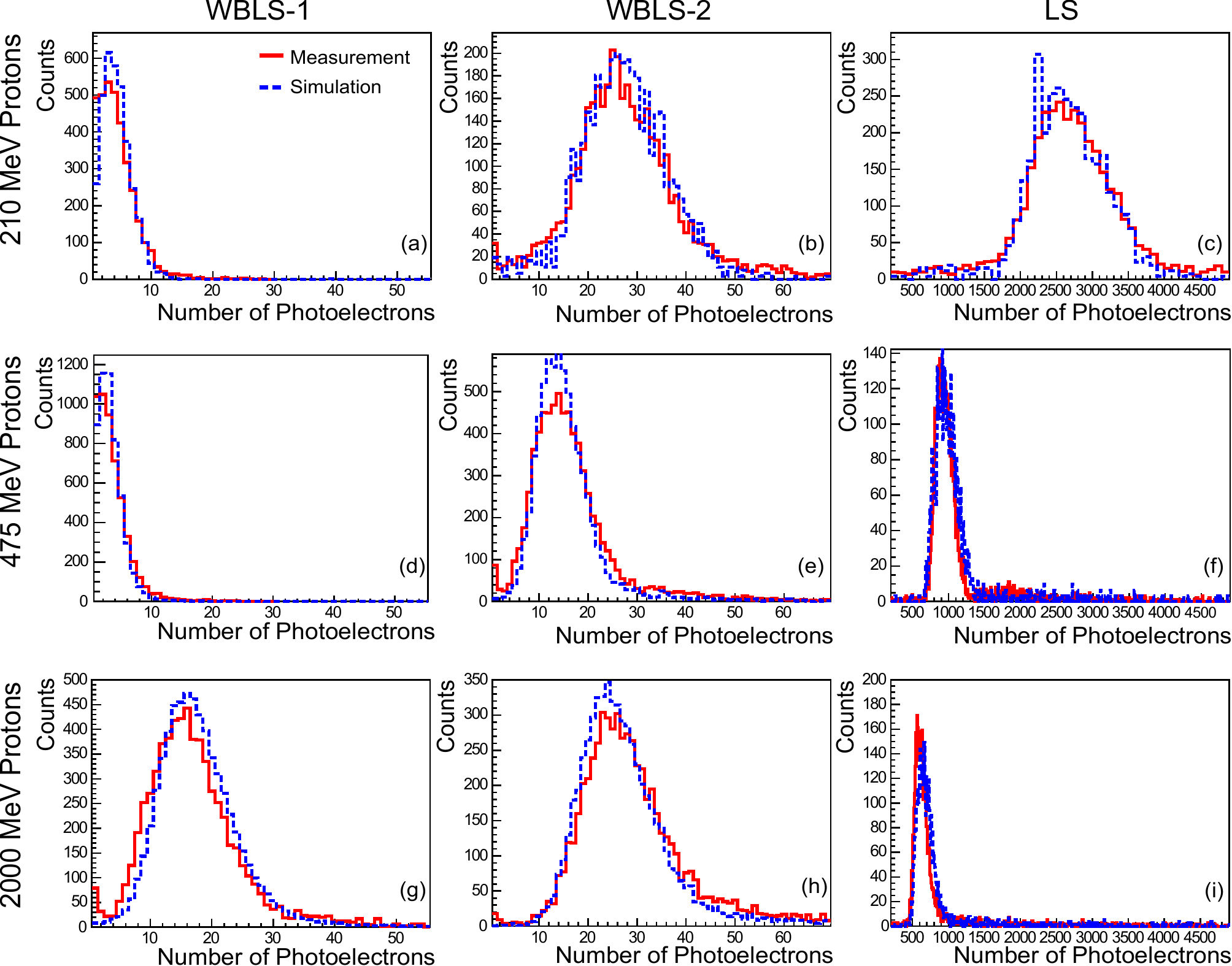}
\caption{Measured (red traces) and simulated (blue traces) photoelectron distributions for the scintillator beam tests in Detector B, Tub 2.}
\label{fig:DetBResults}
\end{figure*}

\begin{figure}[tbp]
\centering
\includegraphics[width=0.8\columnwidth]{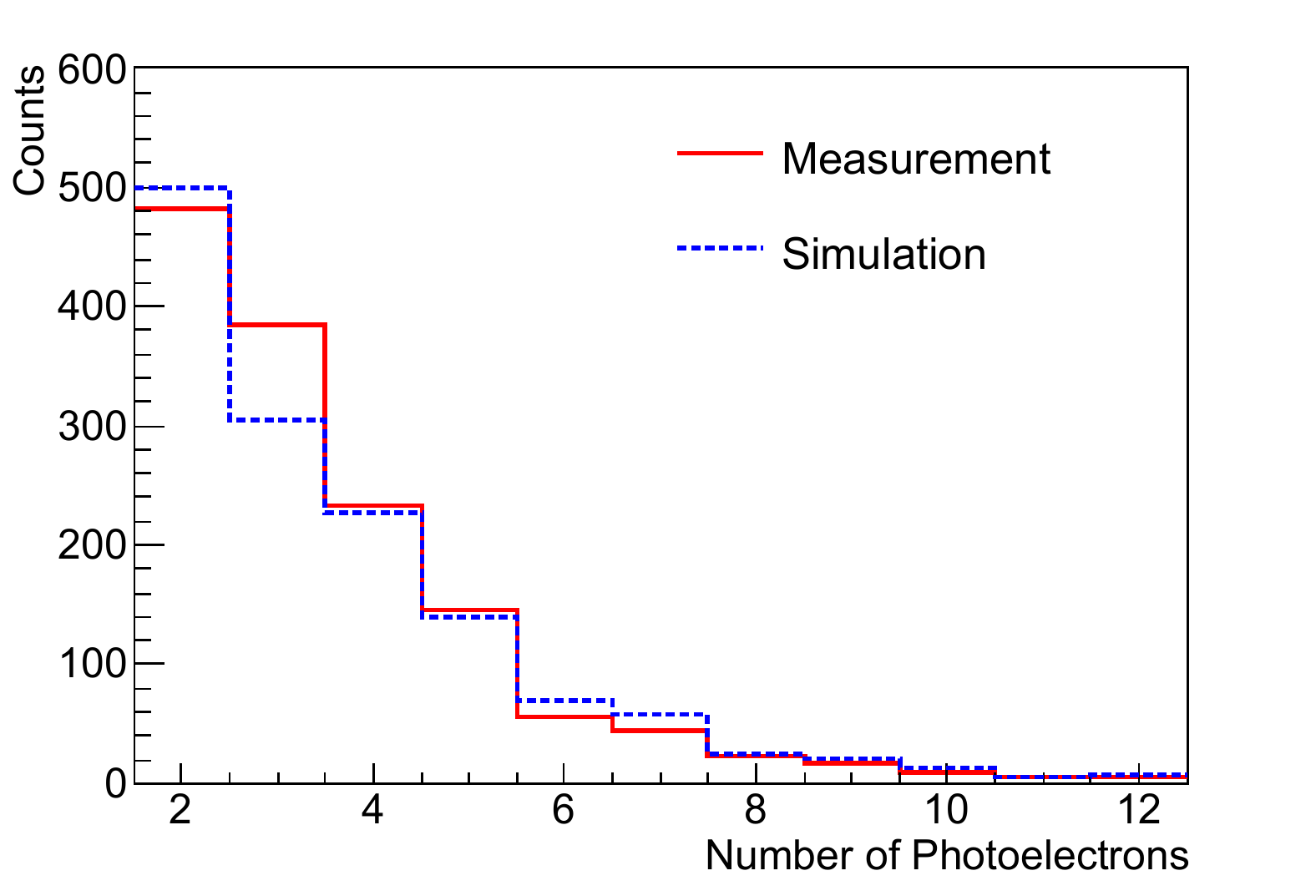}
\caption{Measured (red trace) and simulated (blue trace) photoelectron distribution for a 2 GeV proton beam incident upon water in Detector B, Tub 2. The range over which the minimisation to the measured data was performed is shown.}
\label{fig:DetBWater}
\end{figure}

\section{Results}\label{sec:Results}
\subsection{Detector B}
The simulated and measured photoelectron (PE) distributions are given in figures \ref{fig:DetBResults} and \ref{fig:DetBWater}. Cuts were applied to the hodoscope pulse amplitude and timing, as well as the relative PMT-hodoscope timing to select single proton events from the incident beam. The cuts effectively removed the pedestal noise and did not bias the photoelectron distribution. The measurement of 475 MeV and 210 MeV protons on a water target are not shown as few events were observed during this measurement, and those events that were registered were of very low amplitude.
We attribute the events observed in these measurements to the detection of Cherenkov emission from secondary electrons. The mean number of observed photoelectrons for all measurements and simulations are given in table \ref{tab:MeansDetB}. In addition to the statistical uncertainty arising from the $\chi^2$ analysis of the simulated data, we estimate that there is an additional systematic uncertainty component of $\approx$ 10\% with the main contributions due to uncertainties in the simulation model's input data (such as the refractive index, PMT quantum efficiency, and re-emission probability), the simplified treatment of optical photon scattering, and the effect of two-proton events (Section \ref{sec:BeamInfo}). This systematic component dominates the uncertainty of all simulated results.

\begin{table*}[tbp]
\caption{The most probable energy deposit and mean number of photoelectrons (N$_{PE}$) in Detector B, Tub 2. The most probable energy deposit is taken from the Monte Carlo simulation. We estimate a systematic uncertainty of $\approx$ 10\% for the simulated values. Where uncertainties are given as 0.0, the actual uncertainty is smaller than the smallest reported significant figure.}
\smallskip
\centering
\begin{tabular}{|ccx{3cm}cc|}
\hline
Sample  & Proton Beam Energy    & Most Probable Energy Deposit      & N$_{PE}$, Measured            & N$_{PE}$, Simulated     \\
\hline
Water   & 210 MeV                   & 106.4 $\pm$ 0.3 MeV               & 1.1 $\pm$ 0.0                     & 1.5   $\pm$ 0.2 \\
Water   & 475 MeV                   & 42.0 $\pm$ 0.2 MeV                & 1.7 $\pm$ 0.0                     & 1.3   $\pm$ 0.1 \\
Water   & 2000 MeV                  & 26.6 $\pm$ 0.3 MeV                & 2.8 $\pm$ 0.1                     & 2.5   $\pm$ 0.3 \\
WbLS-1  & 210 MeV                   & 113.9 $\pm$ 0.3 MeV               & 4.5 $\pm$ 0.1                     & 4.4   $\pm$ 0.4 \\
WbLS-1  & 475 MeV                   & 42.2 $\pm$ 0.1 MeV                & 3.6 $\pm$ 0.1                     & 3.4   $\pm$ 0.3 \\
WbLS-1  & 2000 MeV                  & 27.4 $\pm$ 0.2 MeV                & 17.0 $\pm$ 0.2                    & 18.6  $\pm$ 1.9 \\
WbLS-2  & 210 MeV                   & 113.6 $\pm$ 0.3 MeV               & 27.6 $\pm$ 0.3                    & 27.3  $\pm$ 2.7 \\
WbLS-2  & 475 MeV                   & 42.1 $\pm$ 0.2 MeV                & 16.7 $\pm$ 0.1                    & 15.9  $\pm$ 1.6 \\
WbLS-2  & 2000 MeV                  & 27.5 $\pm$ 0.3 MeV                & 30.8 $\pm$ 0.3                    & 29.2  $\pm$ 2.9 \\
LS      & 210 MeV                   & 96.6 $\pm$ 1.4 MeV                & 2588 $\pm$ 21                     & 2622  $\pm$ 262 \\
LS      & 475 MeV                   & 32.5 $\pm$ 0.6 MeV                & 1111 $\pm$ 10                     & 1105  $\pm$ 110 \\
LS      & 2000 MeV                  & 20.9 $\pm$ 0.4 MeV                & 872  $\pm$ 15                     & 933   $\pm$ 93  \\
\hline
\end{tabular}
\label{tab:MeansDetB}
\end{table*}

For Tub 2, the simulated results are generally in agreement with the measurements.
The optimal values of light yield and ionisation quenching parameter are given in table \ref{tab:OptimalVals}. We have used Birks' semi-empirical ionisation quenching model \cite{Birks1964}, and the quenching parameter refers to the material parameter of that model. The light yield of the LS sample is consistent with the typically measured value of the light yield of a liquid scintillator of $\approx 10^{4}$ photons/MeV. The quenching parameter for the LS sample falls in the lower range of the typically measured values of 0.07-0.2 mm/MeV \cite{Peron1996,Torrisi2000,Broda2002,vonKrosigk2013}. The light yield values for the WbLS samples do not linearly scale with the fraction of scintillator, likely due to differences in fluor concentration and solvent type. Both WbLS samples are more susceptible to ionisation quenching than the LS sample. Indeed, the quenching parameter for both WbLS-1 and WbLS-2 is larger than any previously reported value that could be found by the authors for a liquid scintillator. The cause of these extraordinary values will be investigated in future studies.

The 475 MeV and 210 MeV measurements in WbLS-1 resulted in small numbers of photoelectrons with broadened distributions relative to a Poisson distribution. It is also apparent that some low amplitude events with $\approx$ 1 PE contributed to the distribution, as evidence of these events are seen in some of the other measurements. The distribution broadening can be modeled by convolving the simulated distributions with a Gaussian as outlined in \cite{Balagura2006}. However the broadened distribution obtained using this technique were unable to reproduce the shape of the measured photoelectron distributions for all proton energies. Due to the poor agreement of distribution shapes, $\chi^2$ optimisation was quite sensitive to the selection or exclusion of the 1 PE bin in the analysis. The sensitivity of the light yield and quenching parameter to this effect is included in the statistical uncertainties for the WbLS-1 results reported in table \ref{tab:OptimalVals}.

The Tub 1 results are not presented, as it was not possible to obtain agreement with the simulation model that was consistent with the measured distributions across all beam energies for all materials, for any value of the tub reflectance. While the cause of this inconsistency is unknown, we have identified two possible causes that arise from simplifications in our simulation model. Firstly, there may be some wavelength-dependence to the tub reflectance. If this is the case, the effective reflectance for Cherenkov light is different from the effective reflectance of the scintillation and WLS emission, which may explain the discrepancies seen with different particle energies and samples, that have different proportions of Cherenkov and scintillation/WLS light. Another source of error that may contribute to the discrepancy is the model's treatment of optical attenuation. In the model, the optical attenuation length -- the sum of optical absorption and scattering -- is treated as equivalent to the optical absorption length. Treating scattering as equivalent to absorption is not generally a problem in an unreflective detector with low PMT coverage -- a photon that is scattered is likely to be absorbed at the wall. Therefore we do not expect this mechanism significantly influence the results in Tub 2. However, in a highly reflective detector such as Tub 1 the scattered photons are more probable to be detected, so that the conflation of scattering with absorption is a source of systematic error.
This issue will need to be addressed prior to the application of the model to large detector geometries with greater PMT coverage.

\begin{table*}[tbp]
\caption{The optimal values of light yield and ionisation quenching parameter for each scintillator.}
\smallskip
\centering
\begin{tabular}{|ccc|}
\hline
Material    & Light Yield (photons/MeV) & Quenching Parameter (mm/MeV) \\
\hline
WbLS-1      & 19.9  $\pm$ 1.1 (stat.) $\pm$ 2.0 (sys.)          & 0.70  $\pm$ 0.12 (stat.) $\pm$ 0.07 (sys.)             \\
WbLS-2      & 108.9 $\pm$ 0.8 (stat.) $\pm$ 10.9 (sys.)         & 0.44  $\pm$ 0.01 (stat.) $\pm$ 0.04 (sys.)             \\
LS          & 9156  $\pm$ 42 (stat.) $\pm$ 916 (sys.)           & 0.07  $\pm$ 0.01 (stat.) $\pm$ 0.01 (sys.)             \\
\hline
\end{tabular}
\label{tab:OptimalVals}
\end{table*}

\subsection{Detector A}

\begin{figure*}[tbp]
\centering
\includegraphics[width=\textwidth]{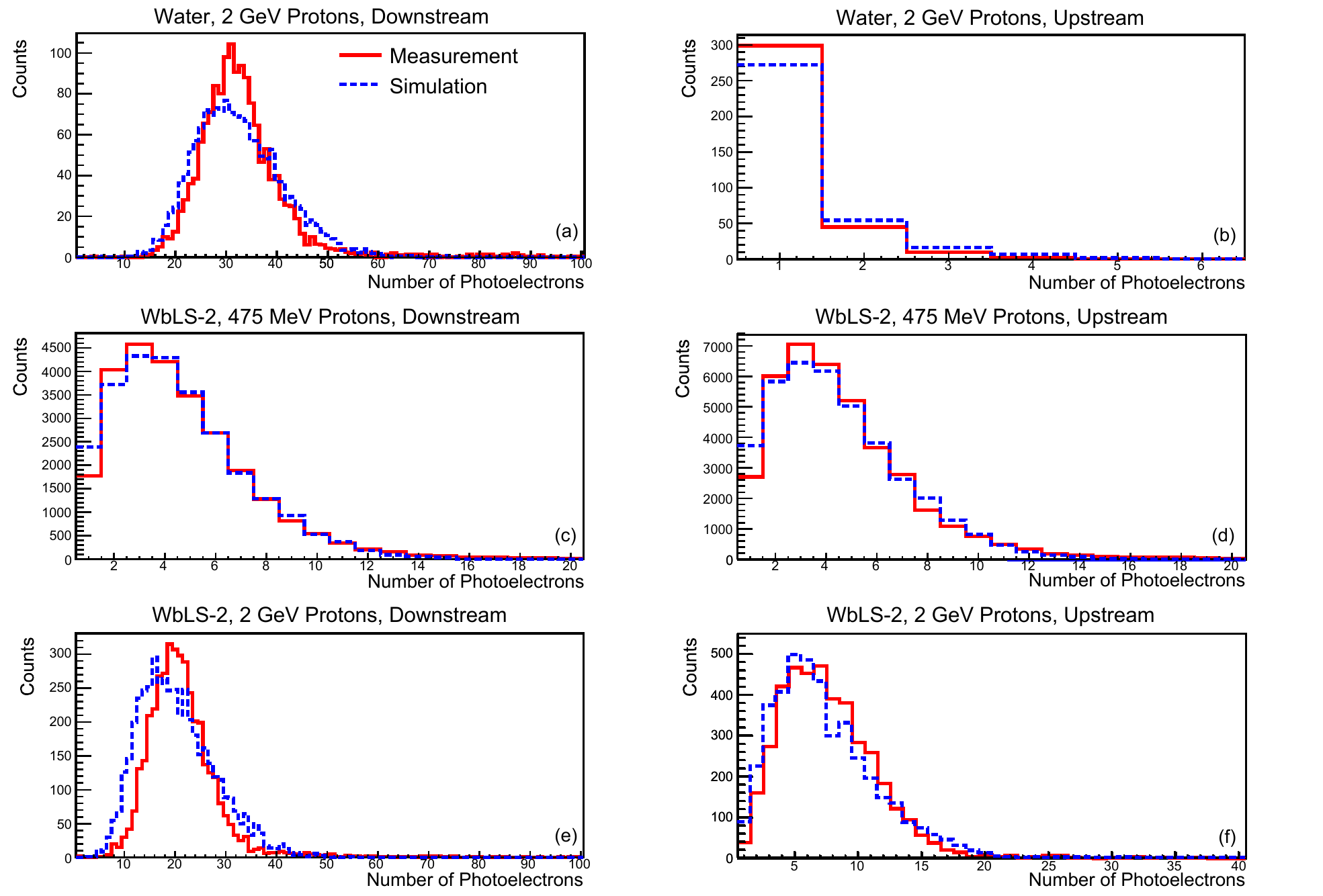}
\caption{Measured (red traces) and simulated (blue traces) photoelectron distributions for the beam tests using Detector A.}
\label{fig:DetAResults}
\end{figure*}

The simulated and measured photoelectron (PE) distributions for Detector A are given in figure \ref{fig:DetAResults}. Quality cuts to the data were applied in the same way as Detector B. The measurement of 475 MeV protons in water are not shown.
The mean number of observed photoelectrons for all measurements and simulations are given in table \ref{tab:MeansDetA}.

\begin{table*}[tbp]
\caption{The mean number of measured and simulated photoelectrons (N$_{PE}$) for Detector A, for photomultipliers located downstream and upstream of the proton beam. We estimate a systematic uncertainty of $\approx$ 10\% for the simulated values. Where uncertainties are given as 0.0, the actual uncertainty is smaller than the smallest reported significant figure.}
\smallskip
\centering
\begin{tabular}{|ccccc|}
\hline
Sample  & Incident Proton Energy    & Photomultiplier   & N$_{PE}$, Measured         & N$_{PE}$, Simulated     \\
\hline
Water   & 475 MeV                   & Downstream        & 1.3 $\pm$ 0.0    & 1.6 $\pm$ 0.2  \\
Water   & 475 MeV                   & Upstream          & 1.2 $\pm$ 0.1    & 1.2 $\pm$ 0.1  \\
Water   & 2000 MeV                  & Downstream        & 33.0 $\pm$ 0.2   & 32.4 $\pm$ 3.2 \\
Water   & 2000 MeV                  & Upstream          & 1.2  $\pm$ 0.0   & 1.4 $\pm$ 0.1  \\
WbLS-2  & 475 MeV                   & Downstream        & 4.7  $\pm$ 0.0   & 4.6 $\pm$ 0.5  \\
WbLS-2  & 475 MeV                   & Upstream          & 4.6  $\pm$ 0.0   & 4.5 $\pm$ 0.5  \\
WbLS-2  & 2000 MeV                  & Downstream        & 21.5 $\pm$ 0.3   & 20.4 $\pm$ 2.0 \\
WbLS-2  & 2000 MeV                  & Upstream          & 7.7  $\pm$ 0.2   & 7.3 $\pm$ 0.7  \\
\hline
\end{tabular}
\label{tab:MeansDetA}
\end{table*}

The simulated results used the optimised values of WbLS-2's light yield and quenching from the Detector B analysis. Whilst fairly good agreement is obtained using these values, the shape of the simulated photoelectron distribution is in general broader than that of the measured distribution. The cause for this behaviour is not fully understood. We have also performed a $\chi^2$ optimisation of the simulated distribution to the measured one by adjusting the simulated light yield using the 475 MeV proton irradiation data. These results suggest an optimal light yield of 109.0 $\pm$ 1.0 
and 110.2 $\pm$ 1.0 
photons per MeV using the downstream and upstream PMTs, respectively. The uncertainties relate only the statistical uncertainty arising from the $\chi^2$ analysis of the simulated data, and we estimate an additional 10\% systematic uncertainty. This good agreement with the Detector B analysis suggests that the model can consistently describe different measurement geometries.

For the 2 GeV proton measurements, many more photons were observed in the downstream PMT than the upstream PMT due to the directional Cherenkov radiation. However, the number of photoelectrons measured in the downstream PMT at 2 GeV for WbLS was substantially less than the number of photoelectrons measured in water at the same energy, which suggests that some proportion of the detectable Cherenkov light is absorbed or scattered by the WbLS. Due to the black detector walls, the upstream detector is largely insensitive to the direct Cherenkov light (as shown in figure \ref{fig:DetAResults}(b)), and therefore this signal can be used to estimate the number of photoelectrons that are measured due to the isotropic light emission processes -- scintillation and WLS.
The absorption probability of the detectable Cherenkov light in the WbLS is given to first order by:
\begin{equation}
P_{abs} =   1-\frac{N_{down}(WbLS, 2~GeV)-N_{up}(WbLS, 2~GeV)}{N_{down}(Water, 2~GeV)}
\end{equation}
where $N_{down}$ and $N_{up}$ are the mean numbers of photoelectrons in the downstream and upstream photomultipliers for a given measurement, respectively.
Our measurements indicate that $P_{abs} \approx 58\%$; that is, 58\% of Cherenkov photons that were measured in the water sample were instead absorbed in the WbLS sample measurement.

The large proportion of Cherenkov photons absorbed or scattered by the ${\mathcal O}(10 {\rm cm})$ path length in WbLS suggests that the optical attenuation length of the WbLS used in this measurement is too short to permit Cherenkov imaging in a very large detector. The optical attenuation length of more recent WbLS formulations has been improved by about 2 orders of magnitude (figure \ref{fig:UVVIS}). To first order, the fraction of absorbed Cherenkov photons simply scales with the optical attenuation length, which suggests that the $\approx 58\%$ loss of Cherenkov photons would occur over a path length of ${\mathcal O}(10 {\rm m})$ using the more recent WbLS formulation; which is a length scale suitable for a large detector. Experimental measurements in a larger detector geometry are needed to more rigorously assess this model prediction.

Our measurements can also give an estimate of the relative proportions of measured WLS and scintillation photons. The mean number of photons detected in the upstream PMT of Detector A for 2 GeV protons incident on WbLS-2 may be written as the sum of the detected scintillation and WLS light components:
\begin{equation}
N_{up}(WbLS, 2~GeV) =   N_{up}^{Scint}(WbLS, 2~GeV) + N_{up}^{WLS(Ckov)}(WbLS, 2~GeV)
\label{eqn:Up2GeV}
\end{equation}
where $N_{up}^{Scint}(WbLS, 2~GeV)$ and $N_{up}^{WLS(Ckov)}(WbLS, 2~GeV)$ are the mean number of detected scintillation photons and WLS Cherenkov photons, respectively.
Any wavelength shifting of scintillation light is disregarded in this approximation.

The mean number of photons detected in the upstream PMT of Detector A for 475 MeV protons incident on WbLS-2 may be written as a similar expression, though without any WLS Cherenkov light component, as 475 MeV protons are at the Cherenkov production threshold:
\begin{equation}
N_{up}(WbLS, 475~MeV) = N_{up}^{Scint}(WbLS, 475~MeV)
\end{equation}

The 2 GeV and 475 MeV protons experience stopping powers in water of 2.0 and 2.8 MeV/cm, respectively. It is possible estimate the difference in scintillation light yield at the different energies using the Birks quenching correction:
\begin{equation}
Q(E) = \frac{\frac{dE}{dx}}{1 + k_{B}\frac{dE}{dx}}
\label{eqn:Birks}
\end{equation}
where $E$ is the proton energy and $k_{B}$ is the quenching parameter. Note that equation \ref{eqn:Birks} assumes a constant stopping power for the particles interacting in the tub. The corrected light yields can then be used to relate the number of scintillation photons measured using the 475 MeV proton beam with the 2 GeV proton beam:
\begin{equation}
N_{up}(WbLS, 2~GeV) = \frac{Q(2 GeV)}{Q(475 MeV)}N_{up}(WbLS, 475~MeV) + N_{up}^{WLS(Ckov)}(WbLS, 2~GeV)
\label{eqn:WLSCkov}
\end{equation}

Equations \ref{eqn:Up2GeV} and \ref{eqn:WLSCkov} permit the calculation of the relative contributions of the scintillation and Cherenkov processes to the measured signal in the upstream PMT. Our results indicate that in the upstream PMT for incident 2 GeV protons; for every detected photon that was produced by scintillation $1.27 \pm 0.05$ photons are detected that are produced by WLS Cherenkov light. This value is consistent with the simulation prediction of 1.28. 
The relatively high proportion of re-emitted Cherenkov light suggests that the detection sensitivity of a WbLS-based detector will be improved significantly beyond what would be expected by the scintillation light yield alone for particles that exceed the Cherenkov threshold in the medium. However, the WLS Cherenkov light is less localised to the position of the energy deposit in the detector and modifies the linearity of the energy response of the scintillator. These complicating factors will need detailed study in a larger measurement geometry in order to evaluate their effect upon track reconstruction and calorimetry.

\section{Summary and Outlook}\label{sec:Conclusion}
The work outlined in this study has made a first measurement of the light production and propagation characteristics of water-based liquid scintillator. Two different WbLS concentrations and pure liquid scintillator were studied. The simulation model developed in this work appears to be fairly robust as it was able to reproduce the measured photoelectron distributions in kilogram-scale detectors for three different proton energies that each probed different light emission properties of the WbLS.
Higher quality WbLS re-emission probability measurements and better treatment of scattering will be used to improve the simulation model. Measurements and model validation in larger test geometries are also required before our model can be used with confidence to predict the performance of a kiloton-scale detector. We have undertaken to develop a 1000 liter scale WbLS detector for this purpose.

\section*{Acknowledgements}
We would like to thank Mike Sivertz, Adam Rusek, and Chiara La Tessa at the NASA Space Radiation Laboratory, as well as Ken Sexton for their assistance with this study. This research was supported by LDRD 12-033 of Brookhaven National Laboratory and by the U.S. Department of Energy, contract number KA2501032.



\end{document}